\documentclass[11pt,a4paper]{article}
\usepackage{jheppub}

\usepackage{amsmath,amssymb}
\usepackage{cancel}
\usepackage{graphicx}
\usepackage{epsfig}
\usepackage{amsmath,amsfonts,amssymb}
\usepackage{multirow}
\usepackage{colortbl}
\usepackage{hhline}
\usepackage{booktabs}
\usepackage{bbm}
\usepackage{bm}
\usepackage{tabulary}
\usepackage{pifont}
\usepackage{bbold}
\usepackage{pifont}

\newcolumntype{K}[1]{>{\centering\arraybackslash}m{#1}}
\title{Mass Unspecific Supervised Tagging (MUST) for boosted jets}

\author[a]{J.~A.~Aguilar--Saavedra}
\author[b]{F. R. Joaquim}
\author[b]{J.~F.~Seabra}
\affiliation[a]{Departamento de F\'{\i}sica Te\'{o}rica y del Cosmos, Universidad de Granada, E-18071 Granada, Spain}
\affiliation[b]{Departamento de F\'{\i}sica and CFTP, Instituto Superior T\'ecnico, Universidade de Lisboa, Av. Rovisco Pais,1, 1049-001 Lisboa, Portugal}

\abstract{Jet identification tools are crucial for new physics searches at the LHC and at future colliders. We introduce the concept of Mass Unspecific Supervised Tagging (MUST) which relies on considering both jet mass and transverse momentum varying over wide ranges as input variables -- together with jet substructure observables -- of a multivariate tool. This approach not only provides a single efficient tagger for arbitrary ranges of jet mass and transverse momentum,
but also an optimal solution for the mass correlation problem inherent to current taggers. By training neural networks, we build MUST-inspired generic and multi-pronged jet taggers which, when tested with various new physics signals, clearly outperform the variables commonly used by experiments to discriminate signal from background. These taggers are also efficient to spot signals for which they have not been trained. Taggers can also be built to determine, with a high degree of confidence, the {\it prongness}  of a jet, which would be of utmost importance in case a new physics signal is discovered.}

\begin{document}

\maketitle
\flushbottom

\section{Introduction}
The high-energy frontier of particle physics has been and will continue to be explored in the decades to come at the Large Hadron Collider (LHC), a machine designed to unveil the intricate dynamics of the Standard Model (SM) and search for new physics signals. Being a proton-proton collider, the LHC abundantly produces sprays of hadronised quarks and gluons (jets), stemming mainly from pure Quantum Chromodynamics (QCD) processes. When sufficiently boosted, the hadronic decay products of SM particles like the $W$, $Z$ and Higgs bosons and the top quark become highly collimated yielding single `fat' jets. This could also happen for new particles decaying hadronically. Actually, multi-jet signals originated from direct or cascade decays of yet unseen particles are predicted in a plethora of theoretical frameworks beyond the SM, ranging from left-right symmetric models~\cite{Aguilar-Saavedra:2015iew} to scenarios with warped extra dimensions~\cite{Agashe:2016rle,Agashe:2016kfr}. The complexity of the various possible jet topologies, and the importance of their identification, fostered the development of discrimination techniques to distinguish (signal) jets produced in collimated decays of heavy particles, from the QCD ones (background). Those methods have been extensively used, for instance, in searches for new gauge bosons, scalar and spin-2 particles~\cite{Sirunyan:2017ukk,Sirunyan:2017hsb,Aaboud:2018juj,Aaboud:2018zba,Aaboud:2018fgi,Sirunyan:2018ikr,Sirunyan:2019vxa,Sirunyan:2019jbg,Aaboud:2018eoy,Aad:2020ddw}, vector-like quarks~\cite{Sirunyan:2017ynj,Sirunyan:2018ncp,Aaboud:2018zpr,Sirunyan:2019xeh} and dark matter~\cite{{Sirunyan:2018gdw}}, as well as in SM measurements~\cite{Sirunyan:2017dgc,Sirunyan:2020hwz}. 

Identification of jets requires (i) quantifying their mass $m_J$, usually 
after applying some `grooming'~\cite{Butterworth:2008iy,Ellis:2009me,Krohn:2009th,Larkoski:2014wba} to remove soft collinear radiation, and (ii)
inferring the number of quarks or gluons clustered inside them ({\em prongs}). The latter procedure, commonly known as {\em tagging}, relies on either a single jet substructure variable like a $N$-subjettiness~\cite{Thaler:2010tr} or energy correlation function~\cite{Larkoski:2014gra,Moult:2016cvt}, or on a multivariate method that takes as input a set of those variables~\cite{Datta:2017rhs} or jet images~\cite{Larkoski:2017jix}. Since quark and gluon jet masses arise mostly from soft radiation, which also modifies jet substructure, mass and substructure variables turn out to be correlated.
Their decorrelation is crucial in several experimental searches, as it prevents artificial peaks from appearing in the $m_J$ distribution of the SM background after imposing jet substructure constraints, and provides a way of improving its normalisation by using sidebands. Moreover, mass decorrelation is a must in new physics searches looking for bumps in jet mass spectra. Given the relevance of this matter, several mass decorrelation methods have been proposed~\cite{Dolen:2016kst,Shimmin:2017mfk,Aguilar-Saavedra:2017rzt,Chang:2017kvc}  (see~\cite{Bradshaw:2019ipy} for a comparison of different methods and ~\cite{Dorigo:2020ldg} for a review) and subsequently applied in a variety of experimental analyses~\cite{Aaboud:2018zba,Sirunyan:2017dgc,Sirunyan:2018ikr,Sirunyan:2018gdw,Sirunyan:2019jbg,Sirunyan:2020hwz,Sirunyan:2019vxa}.

Beyond specific tools designed to identify a certain type of signal (e.g. weak or Higgs bosons and top quarks), more generic ones can also be developed. Supervised taggers use Monte Carlo (MC) simulations of two, three and four-pronged jets as signal, and QCD jets as background. Taking a complete set of substructure variables~\cite{Datta:2017rhs} for both types of jets within some range of  $m_J$ and transverse momentum $p_T$, a multivariate tagging tool such as a neural network (NN)~\cite{Aguilar-Saavedra:2017rzt} or a simpler logistic regression~\cite{Aguilar-Saavedra:2020sxp} can be designed, such that the tagger learns to identify multi-pronged jets as well as new physics objects for which it has not been trained. Alternative proposals focus on unsupervised or weakly-supervised methods, trained directly on data rather than on simulation. Broadly, unsupervised tools are able to distinguish multi-pronged jets from background either by training on samples with different signal and background proportions~\cite{Collins:2018epr,Collins:2019jip,Dillon:2019cqt,Dillon:2020quc,Nachman:2020lpy,Andreassen:2020nkr,Khosa:2020qrz}, or by using autoencoders trained on background regions~\cite{Heimel:2018mkt,Farina:2018fyg,Hajer:2018kqm,Blance:2019ibf,Amram:2020ykb,Cheng:2020dal}. Overall, supervised and unsupervised methods have different strengths and weaknesses and, in particular, supervised methods depend on the details of the parton shower simulation (see appendix F of ref.~\cite{Aguilar-Saavedra:2017rzt}). Still, supervised tools would be certainly essential to claim a new physics discovery if a $5\sigma$ excess is found on data ---  extraordinary claims require extraordinary evidence! Moreover, beyond the discovery of a new physics signal, the identification of its origin obviously requires comparison with Monte Carlo predictions.

Mass decorrelation, as implemented so far in supervised generic taggers, has the disadvantage of showing a residual dependence of the results on the $m_J$ and $p_T$ training ranges. This makes the tagger performance to drop when applied to kinematical regions different from the ones used to train it,
as will be explicitly shown later. To overcome this problem, one could think of assembling an array of taggers in a two-dimensional grid of $m_J$ and $p_T$ to cover the whole kinematical region. But this ad-hoc solution, besides being quite complex, could lead to potential problems with boundary effects.

Up to now, classifiers based on jet substructure have either not taken $m_J$ as input variable~\cite{Thaler:2010tr,Larkoski:2014gra,Moult:2016cvt,Datta:2017rhs,Aguilar-Saavedra:2017rzt,Aguilar-Saavedra:2020sxp}, or have fixed it around some value suitable to tag a specific particle (e.g. a top quark~\cite{Macaluso:2018tck}). In contrast, in this paper we place both $m_J$ and $p_T$ on equal footing as compared to substructure observables by considering the former as training inputs varying over wide kinematical ranges.
This novel approach, which we dub as Mass Unspecific Supervised Tagging (MUST), not only removes the dependence of the tagger efficiency on $m_J$ and $p_T$, but also solves the mass correlation problem in the best possible way by preserving the shape of the $m_J$ distribution after applying the tagger. The taggers built upon MUST cover wide ranges of $m_J$ and $p_T$ (in principle, as wide as wanted) with excellent discrimination performances across all those ranges. The nontrivial challenge of such tools is generating signal multi-pronged jets with continuous $m_J$ and $p_T$ distributions. This will be accomplished by means of a dedicated MC generator. A powerful multivariate method, such as the NN used here, is also required to correctly disentangle mass and $p_T$ effects on jet substructure variables from differences between QCD background and the various multi-pronged signals. 

To further demonstrate the potential of MUST-based taggers, we test their performance to identify complex jet topologies from signals for which they have not been trained. In addition, we build a prongness selection tagger which takes as input jet substructure variables, and could be used to identify new physics signals.

\section{Building the generic taggers}

The first step in order to build the supervised taggers is to generate signal and background jets, which is done as follows. QCD jets are generated with {\scshape MadGraph}~\cite{Alwall:2014hca}, in the inclusive process $pp \to jj$. Event samples are generated in 100 GeV bins from $[200,300]$ GeV to $[2.1,2.2]$ TeV. This guarantees coverage of the entire $p_T$ range up to 2.2 TeV (of course, this arbitrarily chosen domain can be extended). Even though within each bin the events mainly populate the lower end of the interval, the bins are narrow enough to provide a smooth $p_T$ dependence. As for jet mass, the $m_J$ distribution for QCD jets is continuous and we select for our analysis the range $[50,250]$ GeV.

The signal generation is quite more demanding and is carried out with a dedicated MC generator. We implement in {\scshape Protos}~\cite{protos} the process $pp \to ZS$, with $Z \to \nu \nu$ and $S$ a scalar, for which we consider the six decay modes
\begin{align}
&  \text{4-pronged (4P):} && S \to u \bar u u \bar u \,,~ S \to b \bar{b} b \bar{b} \, \notag \\
& \text{3-pronged (3P):} && S \to F \,\nu \,; \quad F \to u d d \,,~ F \to u d b \, \notag \\
& \text{2-pronged (2P):} && S \to u \bar u \,,~ S \to b \bar b \,,
\label{ec:MIdata}
\end{align}
to generate multi-pronged jets ($F$ is a colour-singlet fermion). To remain as model-agnostic as possible, the $S$ and $F$ decays are implemented with a flat matrix element, so that the decay weight of the different kinematical configurations only corresponds to the four-, three- or two-body phase space. These signal MC data are dubbed as Model Independent (MI), being its use motivated by the need of sampling phase space without model prejudice~\cite{Aguilar-Saavedra:2017rzt}. Likewise for the background, signal jet samples are generated in 100 GeV $p_T$ bins. To cover different jet masses, the mass of $S$ (and of $F$ for 3-pronged decays) is randomly chosen event by event within the interval $[30,400]$~GeV, and setting an upper limit $M_S \leq p_T R/2$ to ensure that all decay products are contained in a jet of radius $R=0.8$.\footnote{The $M_S$ interval $[30,400]$ GeV chosen for sample generation is enough for $m_J \in [50,250]$ GeV used in the taggers. Should one wish to extend the $m_J$ range, the selected $M_S$ interval can be extended by changing the parameters in the {\scshape Protos} generator.}The parton-level event samples generated with {\scshape MadGraph} and {\scshape Protos} are passed through {\scshape Pythia}~\cite{Sjostrand:2007gs} for hadronisation and {\scshape Delphes}~\cite{deFavereau:2013fsa} for a fast detector simulation, using the CMS card. Jets are reconstructed with {\scshape FastJet}~\cite{Cacciari:2011ma} applying the anti-$k_T$ algorithm~\cite{Cacciari:2008gp} with $R=0.8$, and groomed with Recursive Soft Drop~\cite{Dreyer:2018tjj}. 

The `model-agnostic' signals in (\ref{ec:MIdata}) allow building supervised generic taggers. In this paper we develop
\begin{itemize}
\item a fully-generic tagger {\tt GenT} using the full set of samples as signal; 
\item multi-pronged taggers {\tt GenT$_\text{4P}$}, {\tt GenT$_\text{3P}$}, {\tt GenT$_\text{2P}$}, which only take the four-, three- and two-pronged jets as signal, respectively.
\end{itemize}
Jet substructure is characterised by the set of variables proposed in~\cite{Datta:2017rhs},
 \begin{equation}
 \left\{ \tau_1^{(1/2)}, \tau_1^{(1)}, \tau_1^{(2)}, \dots , \tau_{5}^{(1/2)}, \tau_{5}^{(1)}, \tau_{5}^{(2)}, \tau_{6}^{(1)}, \tau_{6}^{(2)} \right\} \,,
 \label{ec:taulist}
 \end{equation}
computed for ungroomed jets. We have verified that including higher-order $\tau_n^{(\beta)}$ does not improve tagger discrimination.

The training set is obtained by splitting the considered $m_J$ range $[50,250]$ GeV into four 50 GeV bins. For each of the six types of signal jets in (\ref{ec:MIdata}) and  simulated sample (which, as aforementioned, correspond to different 100 GeV slices of parton-level $p_T$)
we extract $N_0 = 5000$ events from each of the four $m_J$ bins. In the lower $p_T$ samples we drop the higher mass bins, considering the full $m_J$ range only for the $p_T$ bins above 800 GeV. For the {\tt GenT} tagger we take $6 N_0$ background events from each simulated sample and $m_J$ bin, while for the multi-pronged taggers we take $2 N_0$, in order to train the NNs with a balanced sample. We have also explored the possibility of using unbalanced samples with more background than signal events, but we find no improvement in the discrimination power.
In total, the {\tt GenT} and multi-pronged taggers contain $N=4.14\times 10^6$ and $N/3$ events, respectively. The validation sets used to monitor the NN performances are similar to the training ones.

 As anticipated above, we follow a novel approach to train the NNs by considering $m_J$ and $p_T$, varying over a very wide range, together with the 17 substructure observables as inputs. By means of a PCA, 
 we verified that the number of physically relevant combinations is actually smaller; however, since the computational speed is not jeopardised, we keep the full input set. A standardisation of the 19 inputs, based on the SM background distributions, is performed. The NN for {\tt GenT} contains two hidden layers of 2048 and 128 nodes, with Rectified Linear Unit (ReLU) activation for the hidden layers and a sigmoid function for the output one. The NN optimisation relies on the binary cross-entropy loss function, using the Adam~\cite{Adam} optimiser (other generalised loss functions such as the one proposed in~\cite{Murphy:2019utt} do not lead to appreciable improvements). The NNs for the multi-pronged taggers are similar but with hidden layers of 1024 and 64 nodes. We have found no relevant performance improvements of either tagger when using more hidden layers or layers with more nodes.

\section{Tagger performance on multi-pronged jets}
\label{sec:3}

Our taggers are first tested with a variety of multi-pronged jet signals from $W$ bosons, top quarks and new scalars of various masses. (The performance for other types of jets containing hard leptons or photons in addition to quarks is explored in section~\ref{sec:5}.) Namely, 
\begin{itemize}
\item[(i)] two-pronged, with $W \to q \bar q$; new scalars $A \to b \bar b$, with $M_A = 80,\,200$ GeV and $A \to u \bar u$ with $M_A = 200$ GeV;
\item[(ii)] three-pronged, with $t \to W b \to q \bar q b$; 
\item[(iii)] four-pronged, with new scalars $S \to AA \to b \bar b b \bar b$  with $(M_S,M_A) = (80,30),\,(200,80)$ GeV; $S \to WW \to q \bar q q \bar q$ with $M_S = 200$ GeV; $S \to AA \to u \bar u u \bar u$ with $M_S = 80$ GeV, $M_A = 30$ GeV~\cite{Aguilar-Saavedra:2017zuc,Aguilar-Saavedra:2019adu}. 
\end{itemize}
These particles are assumed to be produced with a high boost from the decay of a heavy $Z'$ resonance, for which we choose representative masses $M_{Z'} = 1.1,\,2.2,\,3.3$ TeV.\footnote{Parton-level MC samples are available at https://jaguilar.web.cern.ch/jaguilar/multiprong/} As background, we use quark and gluon jets generated in $pp \to Zq$, $pp \to Zg$, with $Z \to \nu \nu$, in a $1:1$ ratio. All these processes are generated with {\scshape MadGraph}, and passed through the simulation and reconstruction chain described above. The tagger performances are evaluated by comparing the efficiencies for signal ($\varepsilon_\text{sig}$) and background ($\varepsilon_\text{bkg}$) within a narrow $m_J$ interval and with a lower cut on $p_T$, so as to isolate the jet substructure discrimination power from that obtained with any other variable, such as $m_J$ and $p_T$. (An upper cut on $p_T$ is not necessary since both signal and background concentrate towards lower $p_T$ values.) In particular, 
\begin{itemize}
\item  For signals with $M_{Z'}= 1.1, \,2.2,\, 3.3$~TeV, we set 
$p_T \geq 0.5,1.0,1.5$ TeV, respectively, for both signal and background.
\item  For $W$ bosons and scalars decaying as $A \to b \bar b$, $S \to b \bar b b \bar b / u \bar u u \bar u$ with $M_{A,S} = 80$ GeV we use $m_J \in [60,100]$ GeV, while for top quarks $m_J \in [150,200]$ GeV.  When considering $A \to b \bar b / u \bar u $, $S \to q \bar q q \bar q$ with $M_{A,S} = 200$ GeV, we take $m_J \in [160,240]$ GeV. 
\end{itemize}
Besides explicitly showing the receiver operating characteristic (ROC) curve for each signal, we use the area under the ROC curve (AUC) in the $(\varepsilon_\text{sig},\varepsilon_\text{bkg})$ plane to quantify the discriminant power with a single quantity. We also include in our plots vertical lines at signal efficiency of 0.5 and horizontal lines at background rejection of 100, in order to facilitate the visualisation of the intersection of the curves at these reference values. We compare our results with those obtained with the commonly used ratios $\tau_{mn} \equiv \tau_m^{(1)} / \tau_n^{(1)}$.  

\begin{figure}[t]
\begin{center}
\begin{tabular}{cc}
\includegraphics[height=5.5cm]{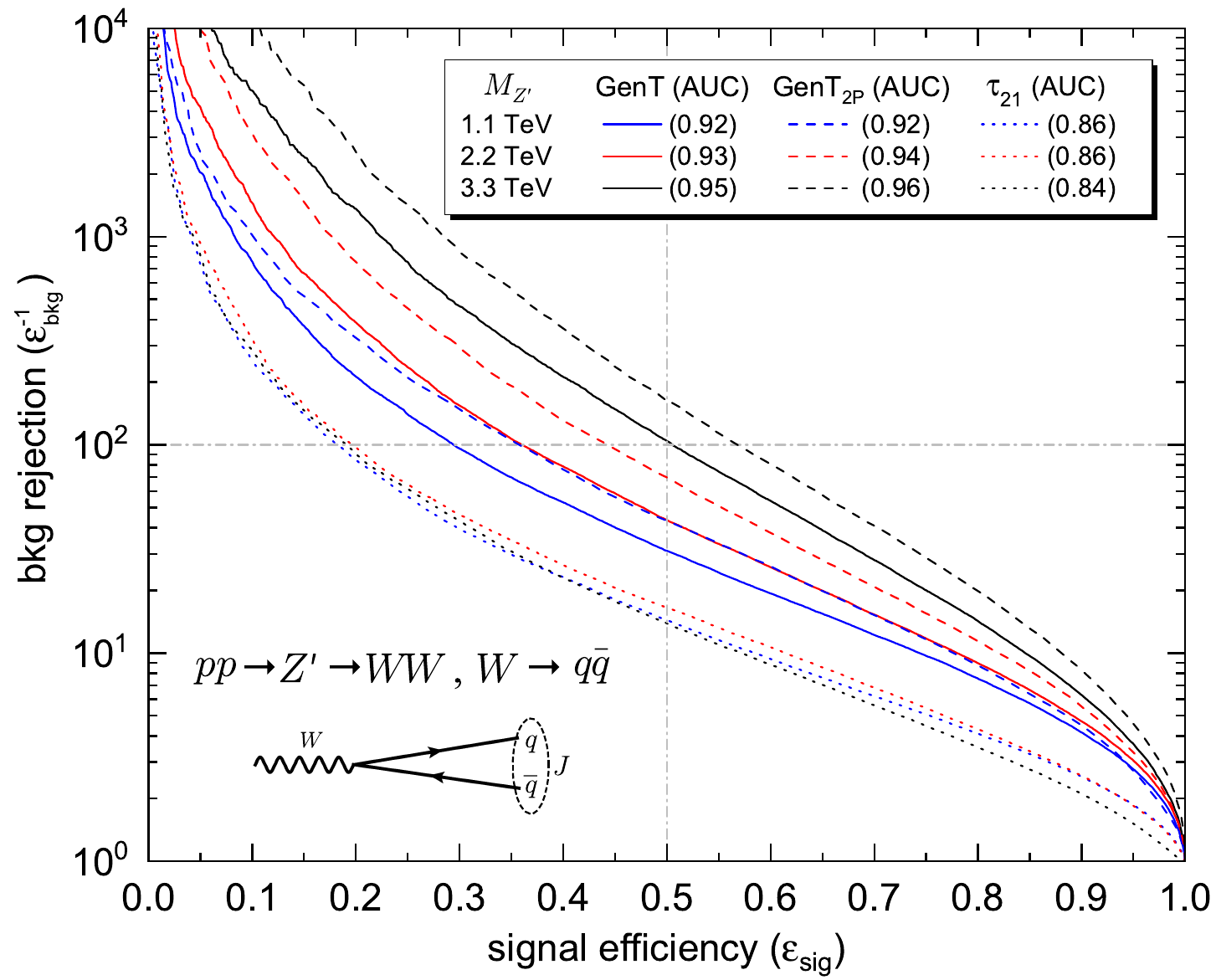} 
& \includegraphics[height=5.5cm]{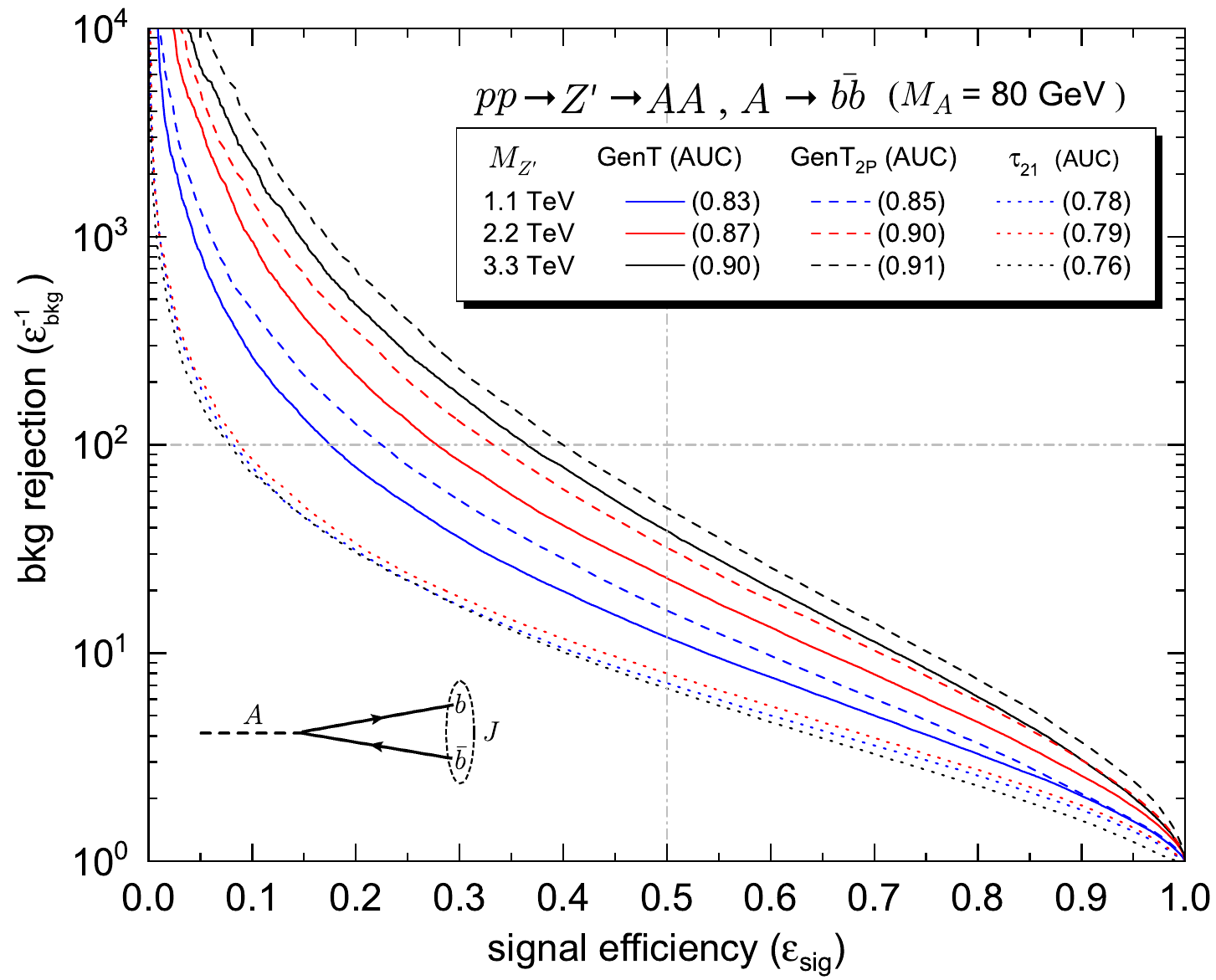} 
\\
\includegraphics[height=5.5cm]{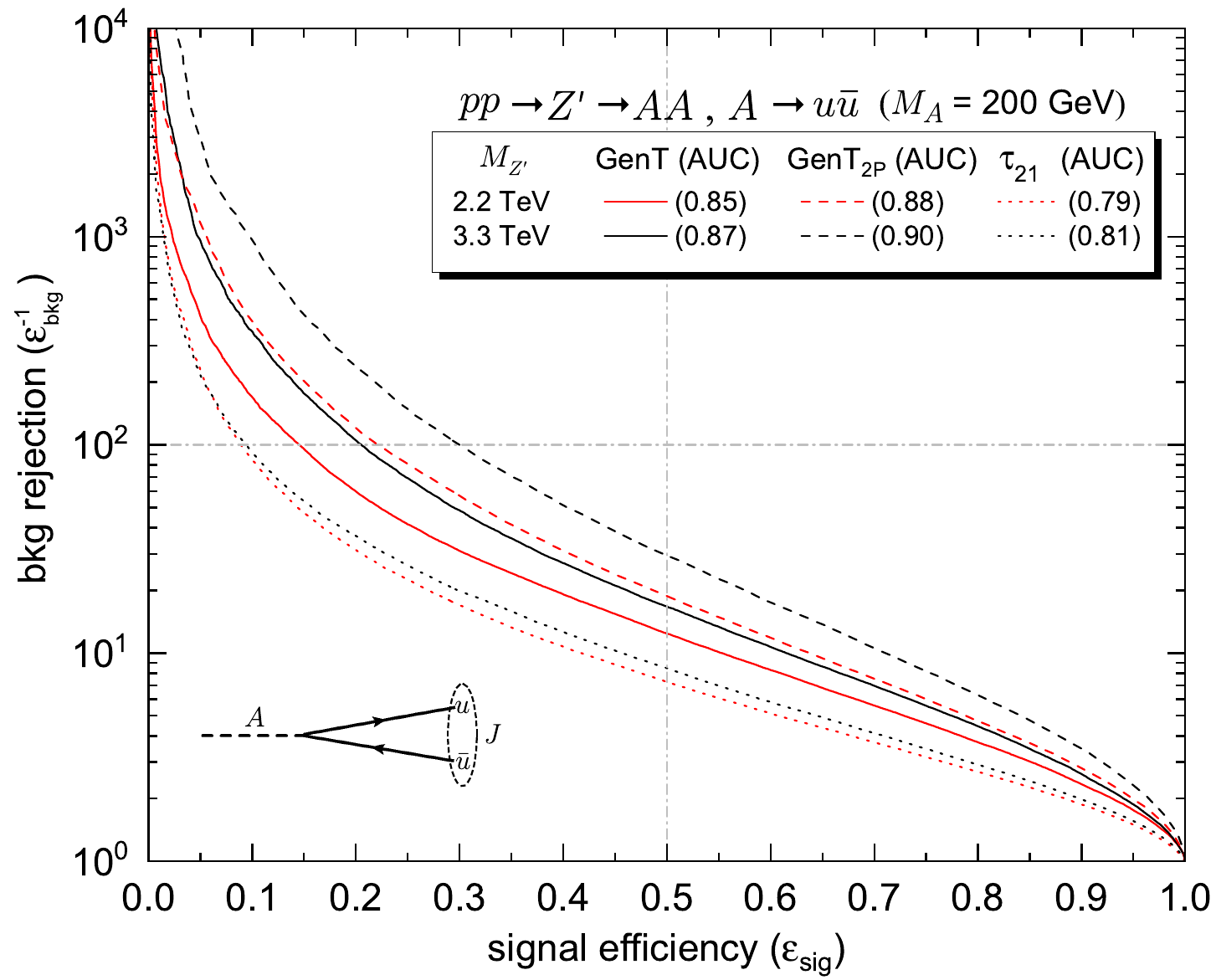} 
& \includegraphics[height=5.5cm]{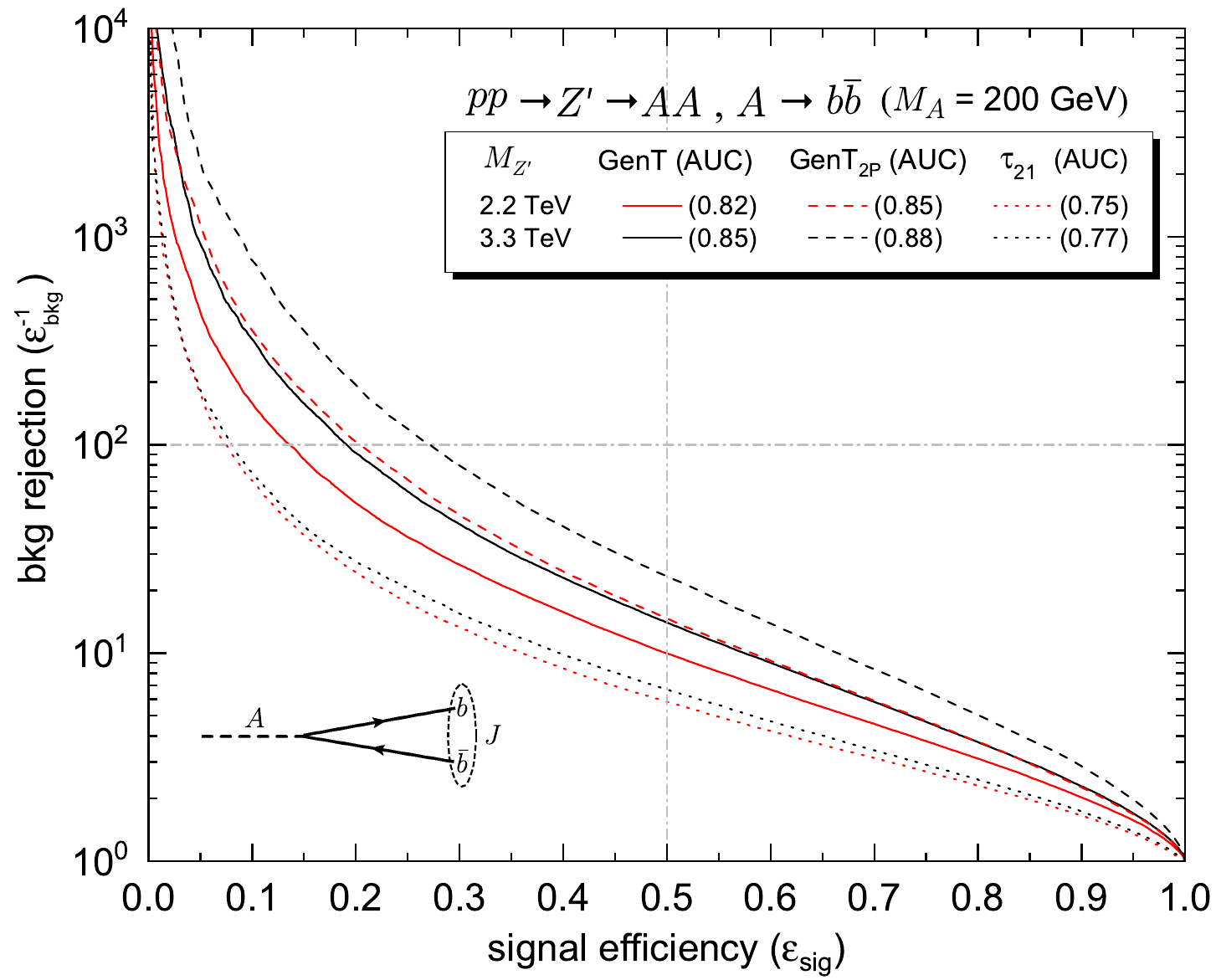} 
\end{tabular}
\caption{ROC curves for two-pronged jet signals (for details see the self-explanatory legends and the main text).}
\label{Fig:2P}
\end{center}
\end{figure}

\begin{figure}[t]
\begin{center}
\includegraphics[height=5.5cm]{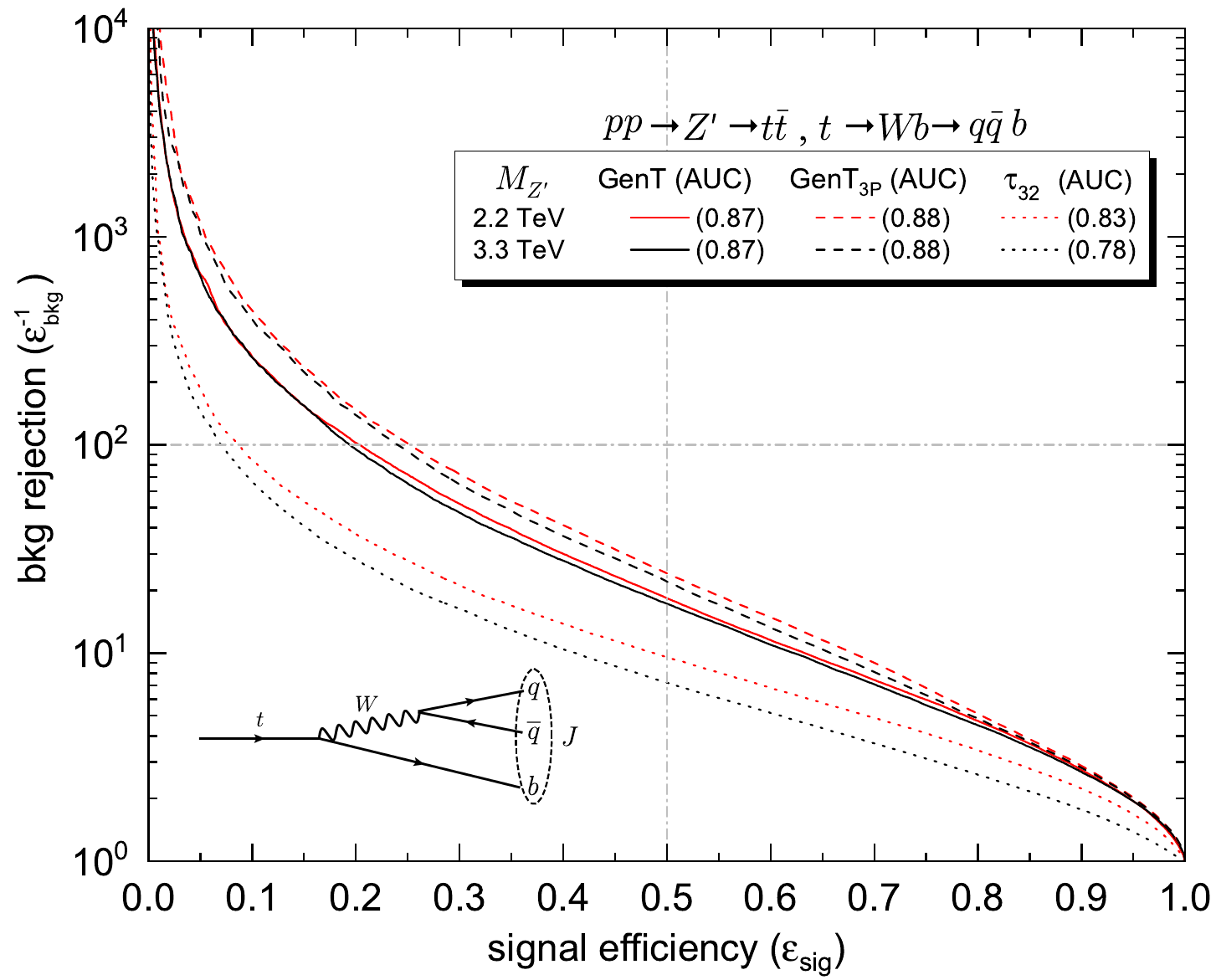} 
\caption{ROC curves for top quark jet signals (for details see the self-explanatory legends and the main text).}
\label{Fig:3P}
\end{center}
\end{figure}

Figure~\ref{Fig:2P} shows the performances of {\tt GenT} and {\tt GenT$_\text{2P}$} for two-pronged jets. We also include the results obtained with $\tau_{21}$, often used as a discriminator by the CMS Collaboration~\cite{Sirunyan:2017ynj,Sirunyan:2019jbg,Sirunyan:2017hsb}. The top-left panel corresponds to $W$ bosons with transverse momenta of 500, 1000 and 1500 GeV. The performance for hadronically-decaying $Z$ bosons is similar. We observe that the performance of {\tt GenT} and {\tt GenT$_\text{2P}$} is remarkable and improves with jet $p_T$, i.e. with increasing $M_{Z'}$, in contrast to $\tau_{21}$. Therefore, our taggers provide an excellent alternative to those used, for example, in diboson resonance searches~\cite{Sirunyan:2019jbg}.

New scalars $A$ decaying into $b \bar b$ are also looked for at the LHC~\cite{Sirunyan:2018ikr,Aaboud:2018eoy}. 
The top-right panel shows the performance for a new scalar $A \to b \bar b$ with $M_S = 80$ GeV. The bottom panels show the results for heavier scalars with $M_S =  200$ GeV; on the left panel we consider decays $A \to u \bar u$ and on the right panel $A \to b \bar b$.  
The performance is very good across all the $m_J$ and $p_T$ range, also improving with $p_T$.

The results for top quarks are shown in Fig.~\ref{Fig:3P} for a couple of $Z'$ masses, and compared with the subjettiness ratio $\tau_{32}$ often used as discriminant~\cite{Sirunyan:2017ukk,Aaboud:2018zpr,Sirunyan:2019xeh,Sirunyan:2018ncp}. {\tt GenT} and {\tt GenT$_\text{3P}$} perform well on top quark jets, although fully-dedicated taggers~\cite{Macaluso:2018tck} perform better (in contrast with ref.~\cite{Macaluso:2018tck}, our ROCs do not include the additional signal discrimination from $m_J$). Still, it is worth noting that generic searches using either of those taggers would not miss top signals, which is precisely the point that we want to verify here.

\begin{figure}[t]
\begin{center}
\begin{tabular}{cc}
\includegraphics[height=5.5cm]{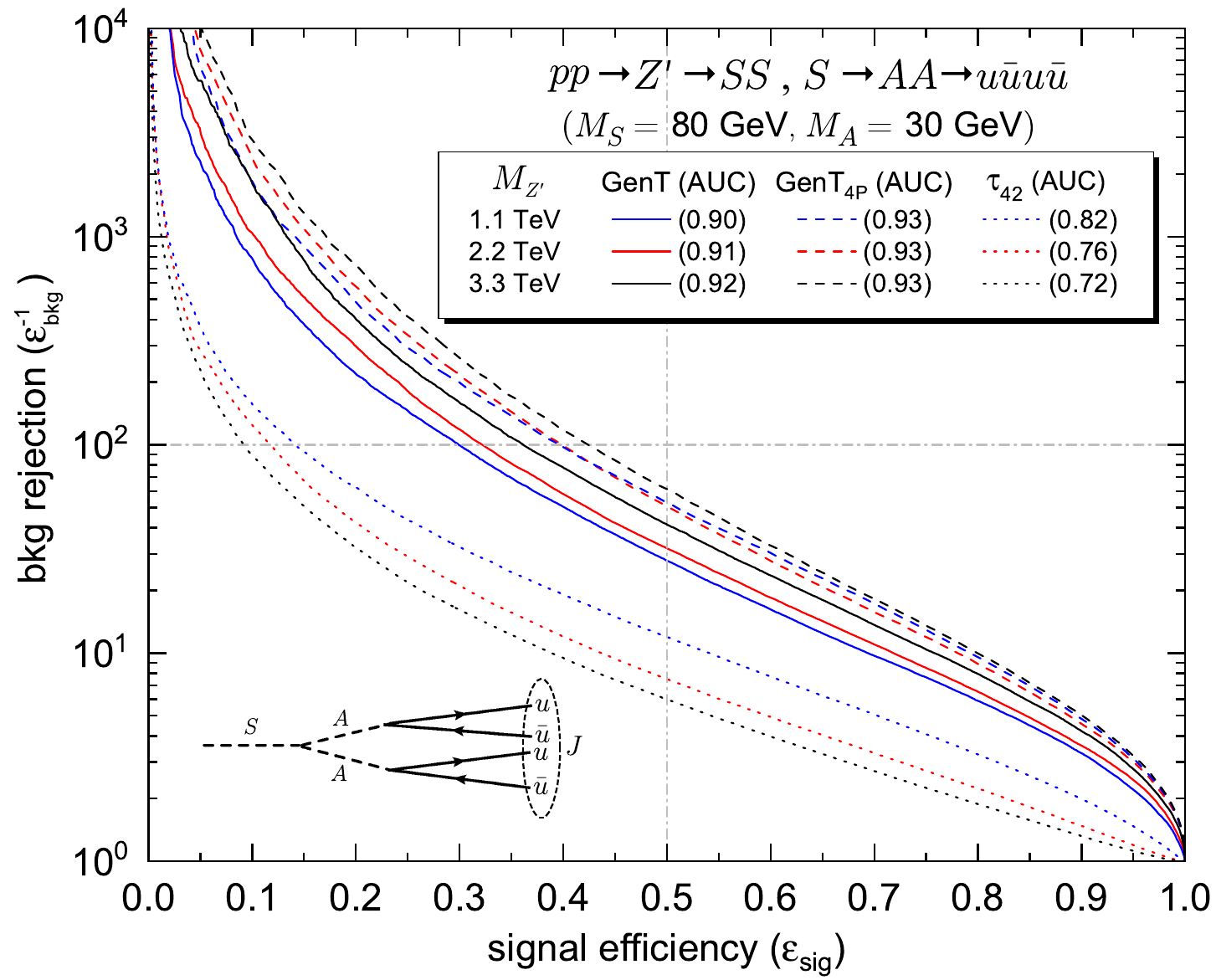} 
& \includegraphics[height=5.5cm]{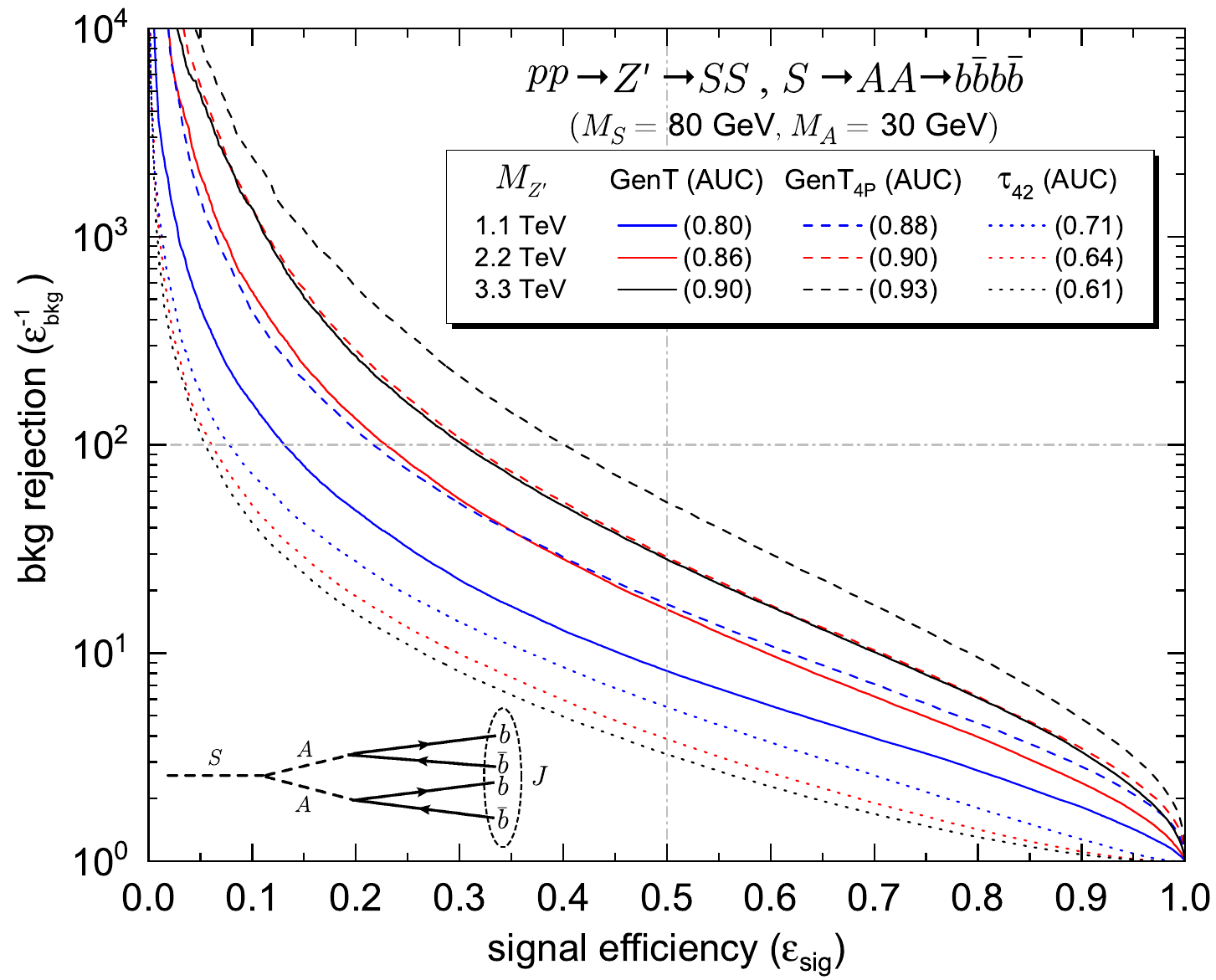} 
\\
\includegraphics[height=5.5cm]{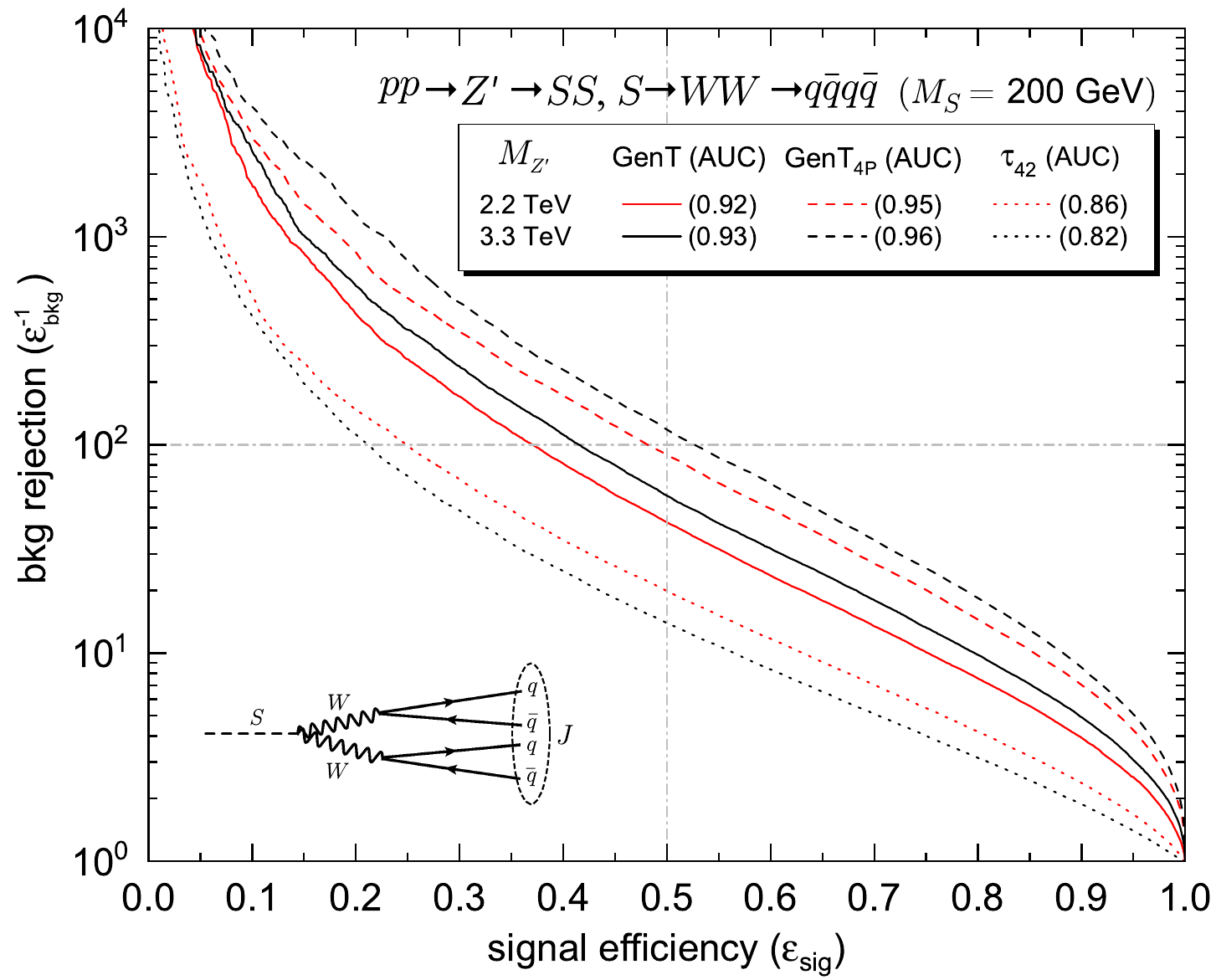} 
& \includegraphics[height=5.5cm]{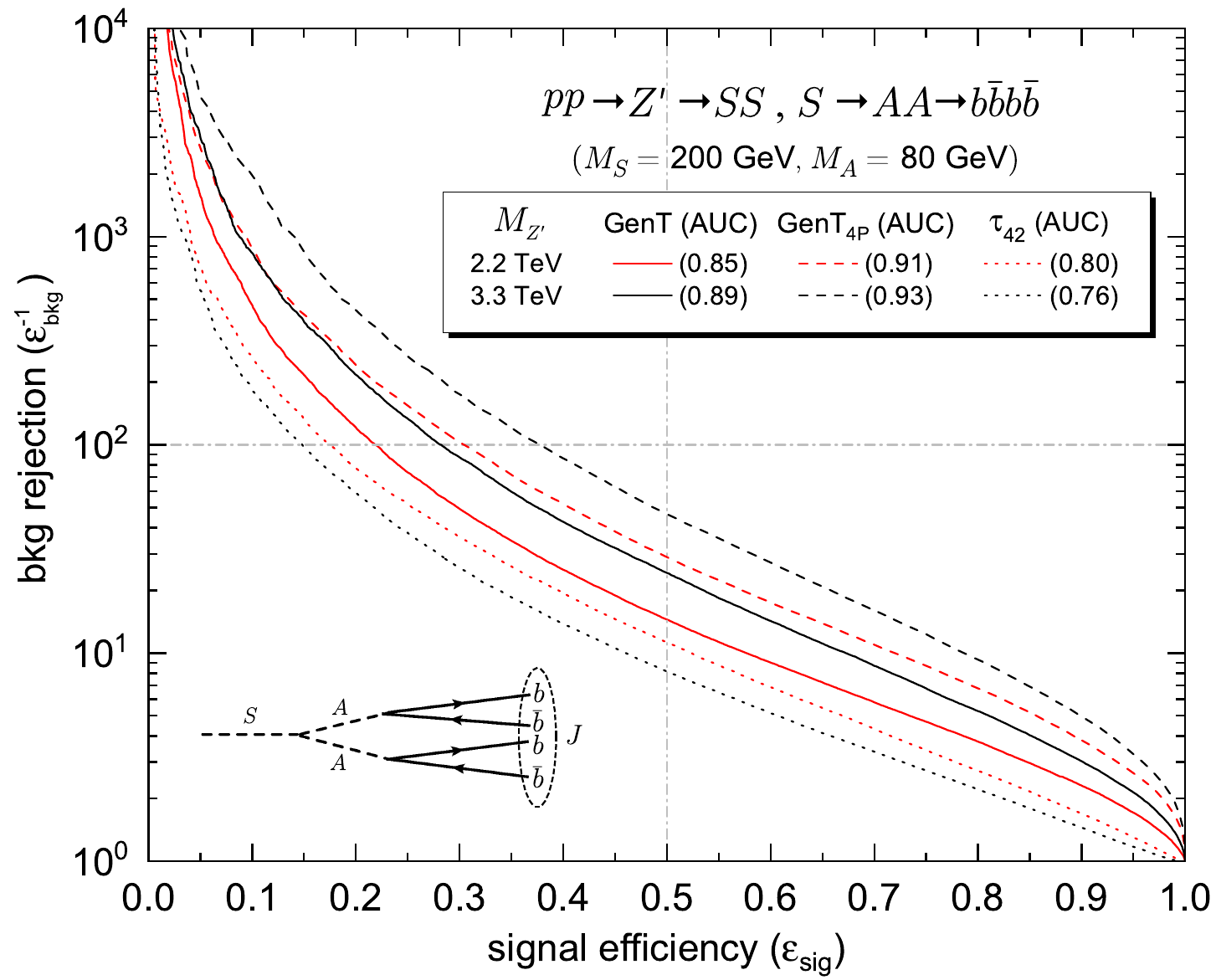} 
\end{tabular}
\caption{ROC curves for four-pronged jet signals (for details see the self-explanatory legends and the main text).}
\label{Fig:4P}
\end{center}
\end{figure}

In Fig.~\ref{Fig:4P} the results for four-pronged jets are shown. For comparison we show $\tau_{42}$, which happens to be the $\tau_{4n}$ with highest AUC. The top panels correspond to signals with $M_S = 80$ GeV, decaying via a pair of lighter scalars $A$ into four $u$ quarks (left) or four $b$ quarks (right). In the bottom panels we present the results for heavier scalars with $M_S = 200$ GeV, either decaying into light quarks via a $WW$ pair (left) or into four $b$ quarks (right) via an $AA$ pair. Again, the {\tt GenT} and {\tt GenT$_\text{4P}$} performances are excellent, exhibiting, for instance, a background rejection better than that of $\tau_{42}$ by a factor of ten, for $\varepsilon_{\rm sig}=0.5$. As expected, in all cases we observe that the multi-pronged taggers provide a higher discrimination power than the generic one for their corresponding multi-pronged signals, but of course they are less general. 

\begin{figure}[t]
\begin{center}
\begin{tabular}{cc}
\includegraphics[height=5.5cm]{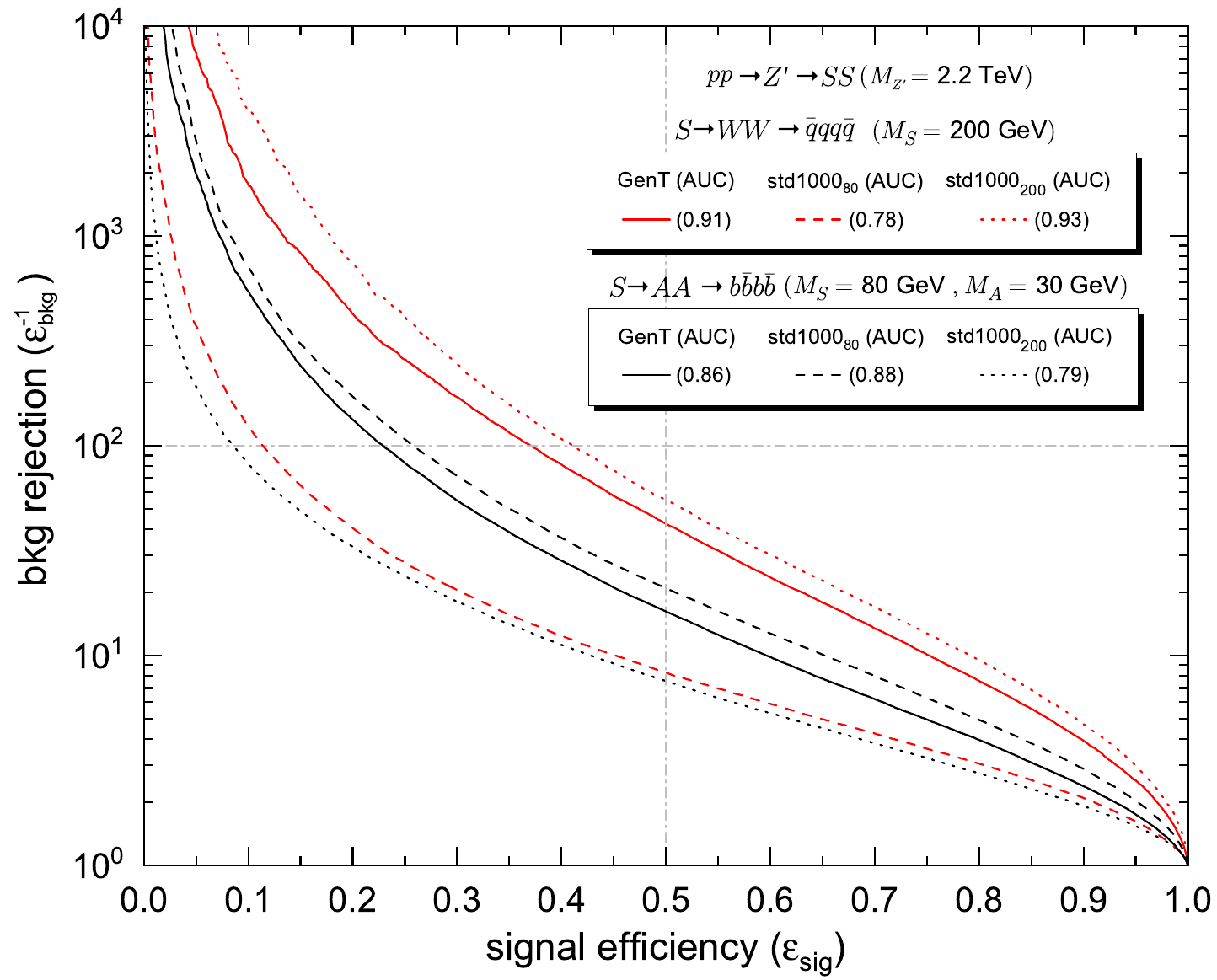} 
& \includegraphics[height=5.5cm]{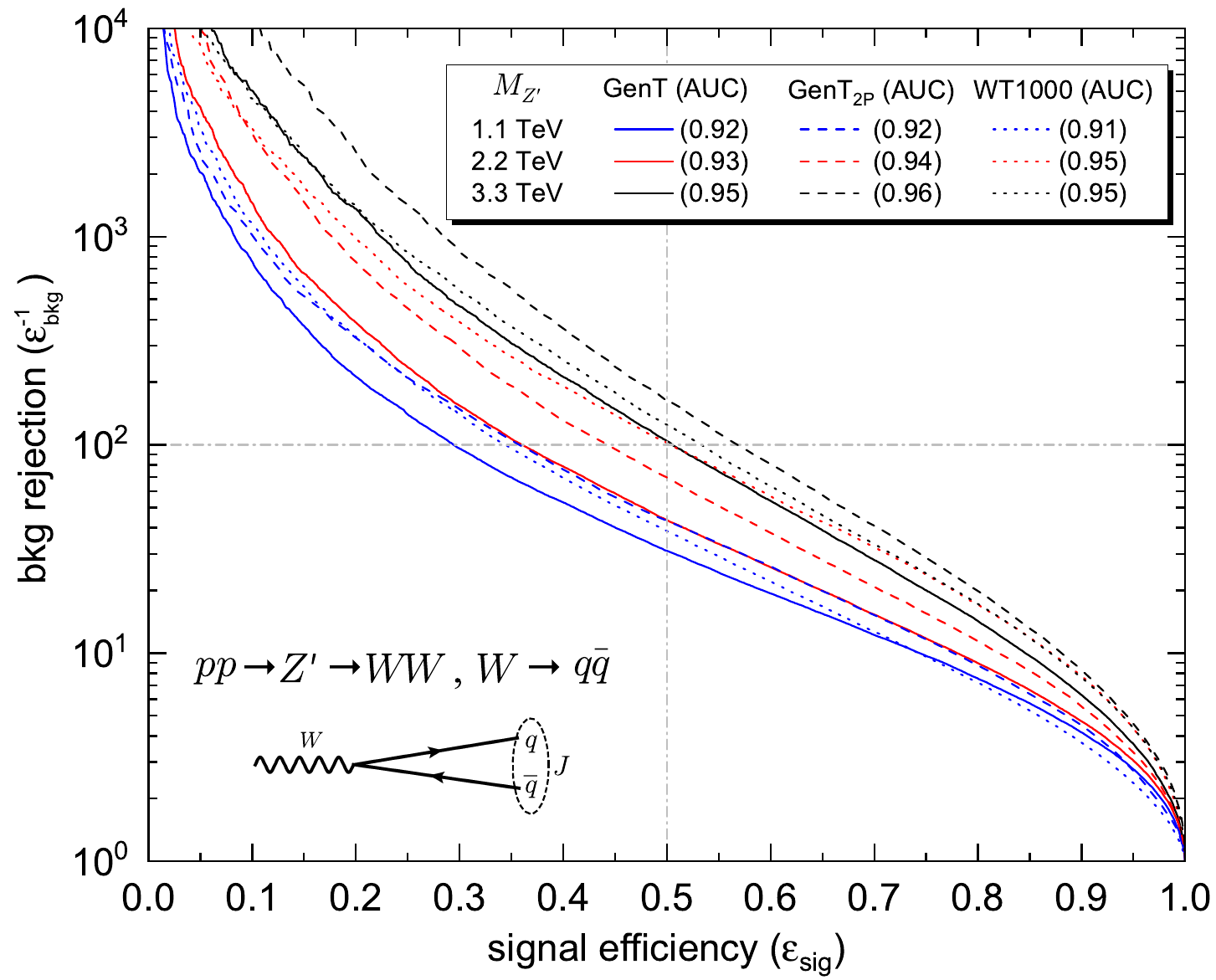} 
\end{tabular}
\caption{Left: comparison of the performances of {\tt GenT} and taggers {\tt pca1000$_{80}$} and {\tt pca1000$_{200}$} trained on narrow jet mass and $p_T$ intervals, for two examples of four-pronged jet signals. Right: comparison of the performances of {\tt GenT}  and {\tt GenT$_\text{2P}$} with a tagger {\tt WT1000} specifically designed for $W$ bosons. }
\label{Fig:pca}
\end{center}
\end{figure}

We are now in position to compare our results with those obtained with taggers trained on narrow $m_J$ and $p_T$ intervals with PCA mass decorrelation~\cite{Aguilar-Saavedra:2017rzt}. We consider signals $S \to AA \to b \bar b b \bar b$ with $M_{Z'} = 2.2$ TeV, $M_S = 80$ GeV and $S \to WW \to q \bar q q \bar q$ with $M_{Z'} = 2.2$ TeV, $M_S = 200$ GeV, as in Fig.~\ref{Fig:4P}. Following~\cite{Aguilar-Saavedra:2017rzt}, we build the PCA-decorrelated taggers {\tt pca1000$_{80}$} and {\tt pca1000$_{200}$} for $p_T \geq$ 1 TeV and
$m_J \in [60,100]$ GeV, $m_J \in [160,240]$ GeV, respectively. The corresponding ROC curves are plotted in Fig.~\ref{Fig:pca} (left). The taggers trained on a narrow mass interval close to the signal mass perform slightly better, but are much less efficient when applied to masses out of the training region.

It is also interesting to compare the performance with a tagger specifically designed for $W$ bosons. We generate a second sample of $Z' \to WW$ with $M_{Z'} = 2.2$ TeV and the $W$ bosons decaying hadronically. We train a PCA-decorrelated tagger {\tt WT1000} using $W$ jets in this sample as signal and QCD jets as background, with $p_T \geq$ 1 TeV and $m_J \in [60,100]$ GeV. The architecture of the NN is the same used for the taggers in Ref.~\cite{Aguilar-Saavedra:2017rzt}. This tagger is then applied to the three $Z' \to WW$ signals with $M_{Z'}= 1.1, \,2.2,\, 3.3$~TeV previously considered. The ROC curves are shown on the right of Fig.~\ref{Fig:pca}, together with the results for {\tt GenT} and {\tt GenT$_\text{2P}$}. From the comparison it is found that, as expected:
\begin{itemize}
\item[(i)] {\tt WT1000}  performs slightly better than {\tt GenT$_\text{2P}$} in the interval $p_T \geq$ 1 TeV and $m_J \in [60,100]$ GeV, in which the former is trained.
\item[(ii)] {\tt WT1000} performs slightly worse than  {\tt GenT$_\text{2P}$} for other momenta $p_T \geq 500$ GeV, $p_T \geq 1.5$ TeV.
\end{itemize}
A $W$ tagger could also be built upon the MUST method to cover $W$ jets with a wide range of $p_T$. But in this case we expect little differences with respect to {\tt GenT$_\text{2P}$}: such $W$ tagger would have a more uniform performance across all $p_T$ ranges, at the expense of some overall degradation, as also observed in the comparisons of Fig.~\ref{Fig:pca} (left).

\section{Mass decorrelation}
\label{sec:4}

\begin{figure}[t]
\begin{center}
\begin{tabular}{cc}
	\includegraphics[width=0.435\textwidth]{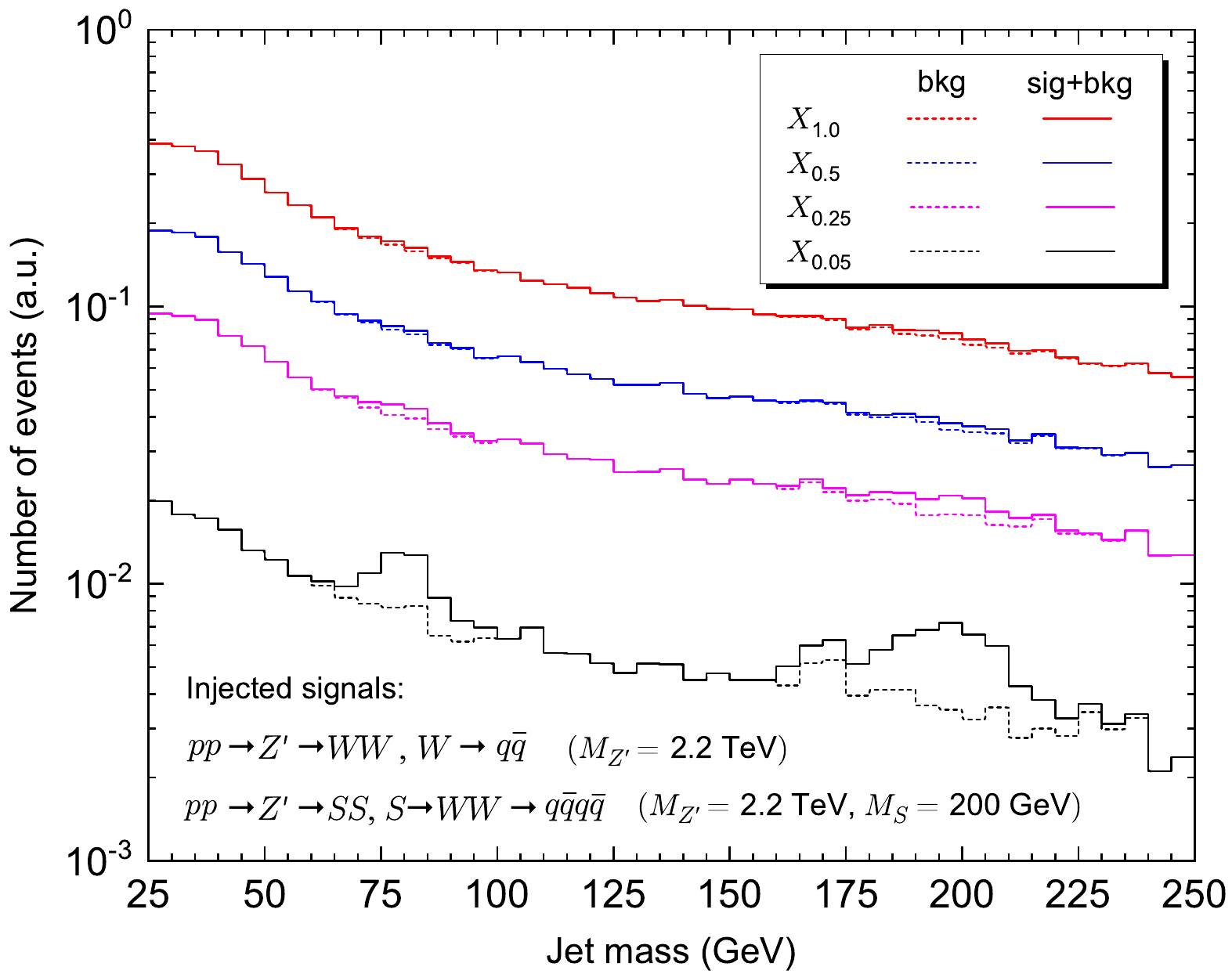} &
	\includegraphics[width=0.435\textwidth]{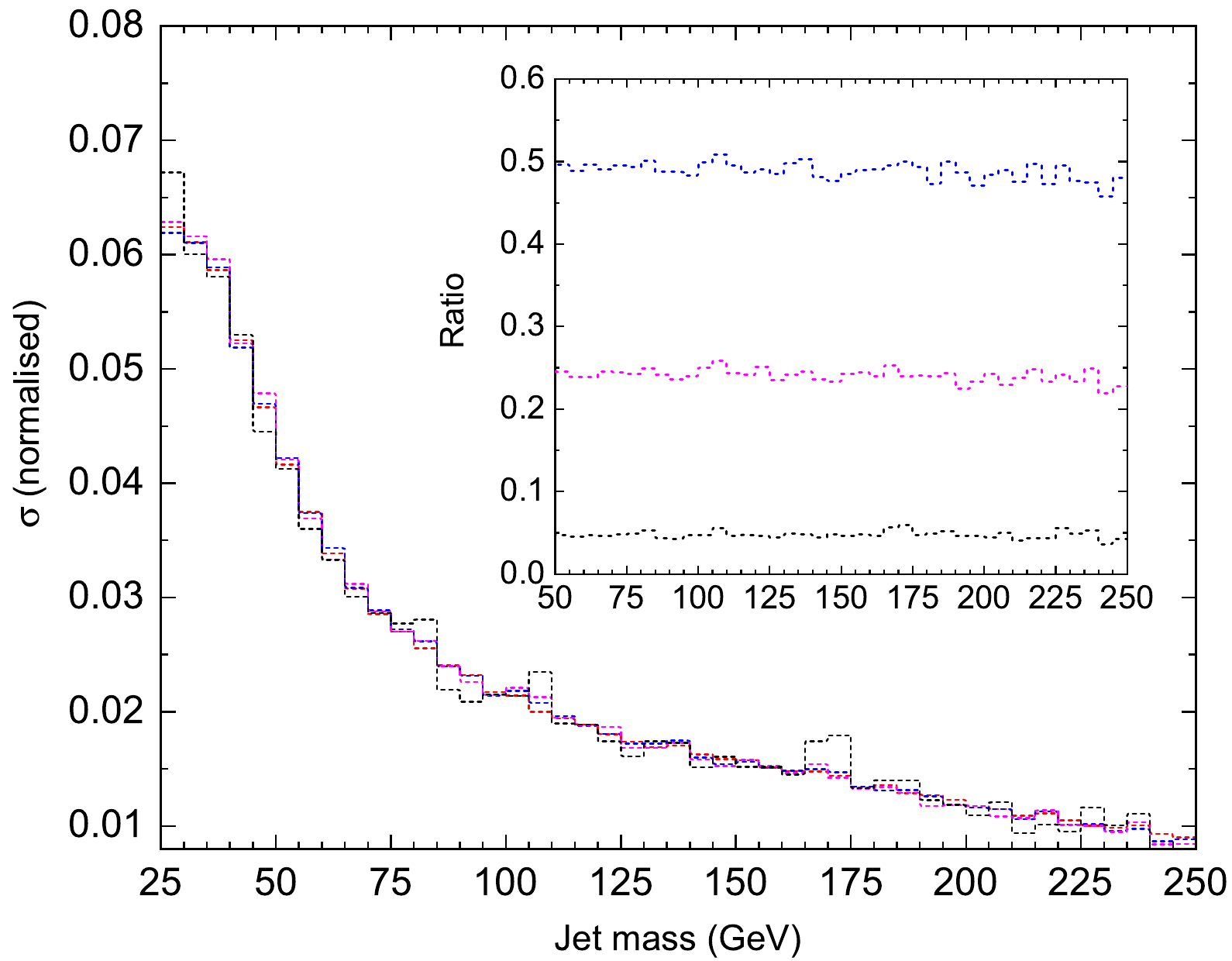}
\end{tabular}
\caption{Left: Jet mass distribution for the SM background plus two injected signals, after the application of increasingly tighter cuts on the NN score. Right: normalised background distributions before and after cuts (main plot), and ratios of distributions after/before cuts (inner plot) -- see the text.}
\label{Fig2}
\end{center}
\end{figure}

Since our taggers are sensitive to multi-pronged jet signals across
the whole $m_J$ and $p_T$ ranges, the SM background shape can be preserved by the simple method of varying the event selection threshold, as done by the CMS Collaboration e.g. in~\cite{Sirunyan:2018ikr}. Let us show this explicitly with an example using {\tt GenT} with a two-pronged ($W$) and a four-pronged ($S$) signal. We define the variable $\rho = 2 \log m_J/p_T$ and consider a two-dimensional grid $(\rho,p_T)$ with $\rho \in [-9,0]$ in bins of width 0.2, and $p_T \in [0.25,2.2]$ TeV in bins of 50 GeV. Within each bin, we compute the $5\%$, $25\%$ and $50\%$ percentiles of the NN score $X$, which we label as $X_{0.05}$, $X_{0.25}$ and $X_{0.5}$, respectively. Fig.~\ref{Fig2} (left) shows the resulting jet-mass distribution of the SM background for $p_T \geq 1$ TeV plus the 
$W$ and $S$ injected signals with $M_{Z'} = 2.2$ TeV, after applying event selections $X \leq X_{0.5},X_{0.25},X_{0.05}$ (the uncut distribution is labelled as $X_{1.0}$). By construction, the varying-threshold scheme keeps the background distribution after selection, and the injected signals show up when the cut is sufficiently tight. The right panel shows the normalised background distributions (main plot) as well as the ratios $X_{0.5}/X_{1.0}$, $X_{0.25}/X_{1.0}$ and $X_{0.05}/X_{1.0}$ (inner plot).
Our generic taggers therefore provide a perfect solution to the mass correlation problem of jet substructure observables.

\section{Tagger performance for other types of jets}
\label{sec:5}

We now address the question of whether the taggers designed to detect multi-pronged jets containing two or more quarks are also able to identify as `signal' other types of complex jets not used in the NN training.\footnote{For non-complex non-SM jets, see e.g. ref.~\cite{Park:2017rfb} our taggers are likely not sensitive.}
Or whether, on the contrary, our taggers classify them as background-like and {\em reduce} the significance of such potential signals. In this respect, it is worth noting that the variable $\tau_{21}$ reduces the significance of signals with four-pronged jets, as pointed out in ref.~\cite{Aguilar-Saavedra:2017zuc} and shown explicitly in Fig.~\ref{Fig:4P}. For our tests we consider
\begin{itemize}
\item[(i)] jets containing two quarks and a hard electron, resulting from a heavy neutrino decay $N \to e q \bar q$ mediated by an off-shell $W'$ boson, with  $M_N = 80\,,~200$ GeV; 
\item[(ii)] jets containing two quarks and two hard photons~\cite{AguilarSaavedra:2020wmg}, resulting from $S \to AA \to b \bar b \gamma \gamma$  with $(M_S,M_A) = (80,30)$ and  $(200,80)$ GeV. 
\end{itemize}
The cuts on jet mass and $p_T$ for the evaluation of the performance are the same described in section~\ref{sec:3}.
\begin{figure}[t]
\begin{center}
\begin{tabular}{cc}
\includegraphics[height=5.5cm]{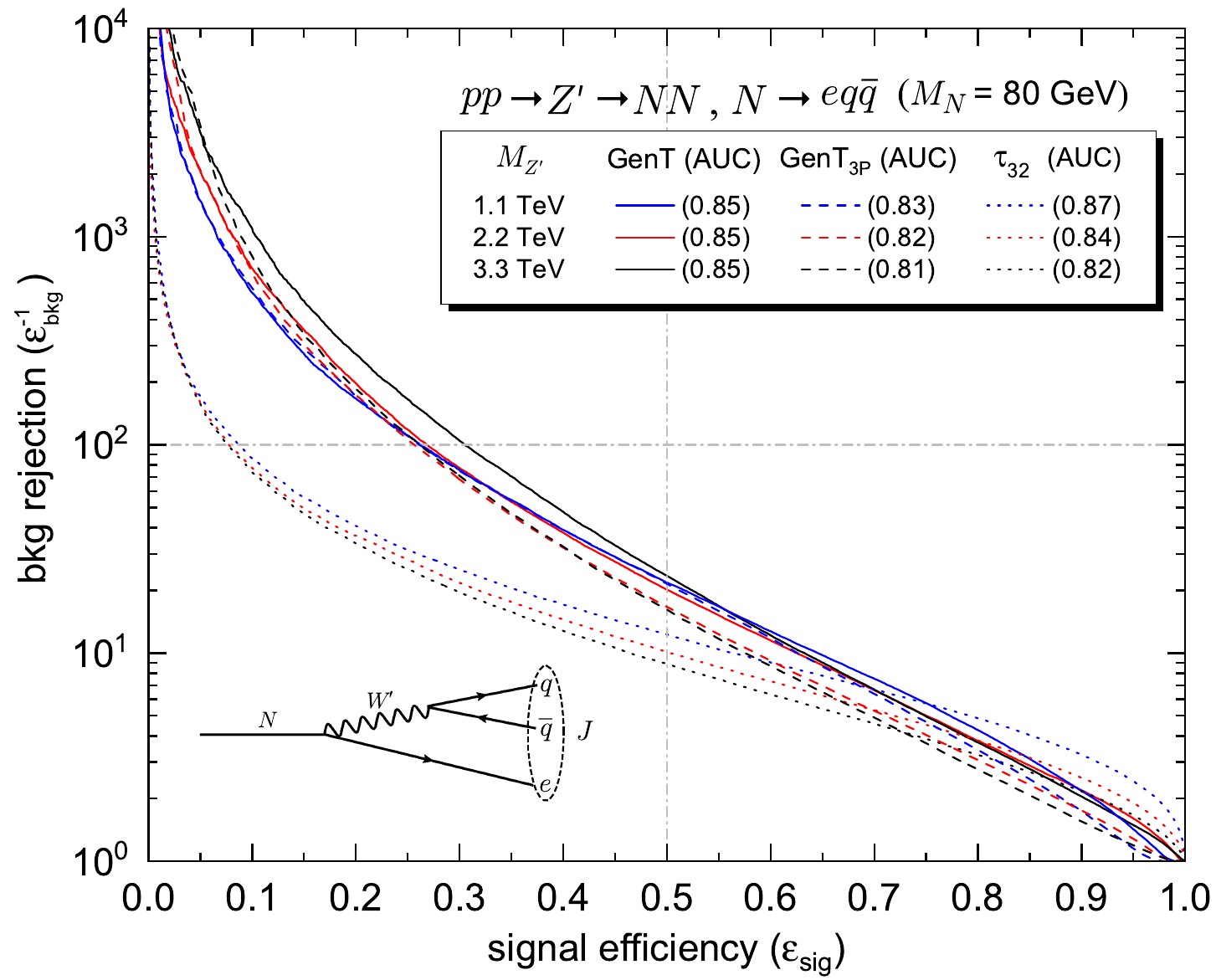} 
 & \includegraphics[height=5.5cm]{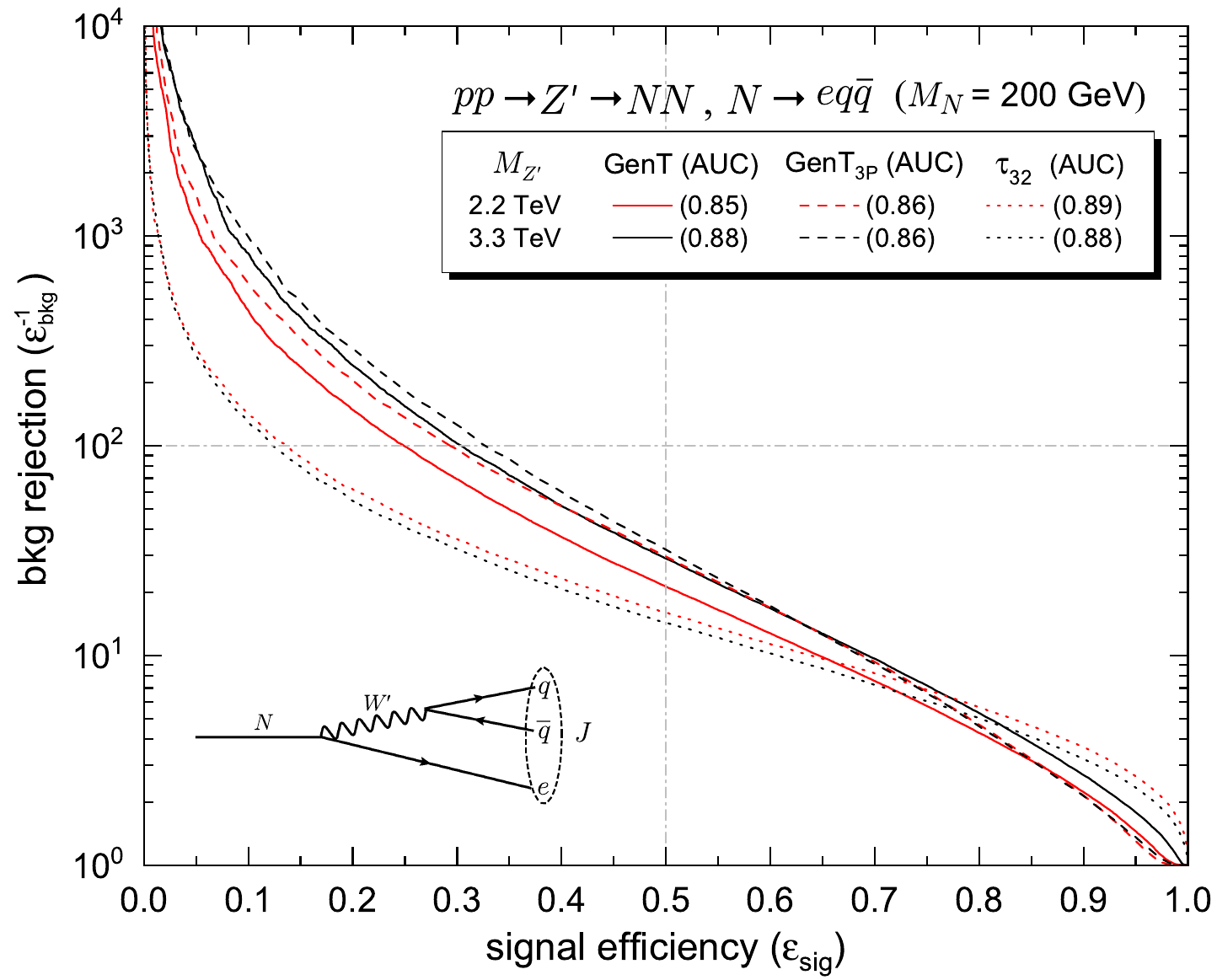} 
\\
\includegraphics[height=5.5cm]{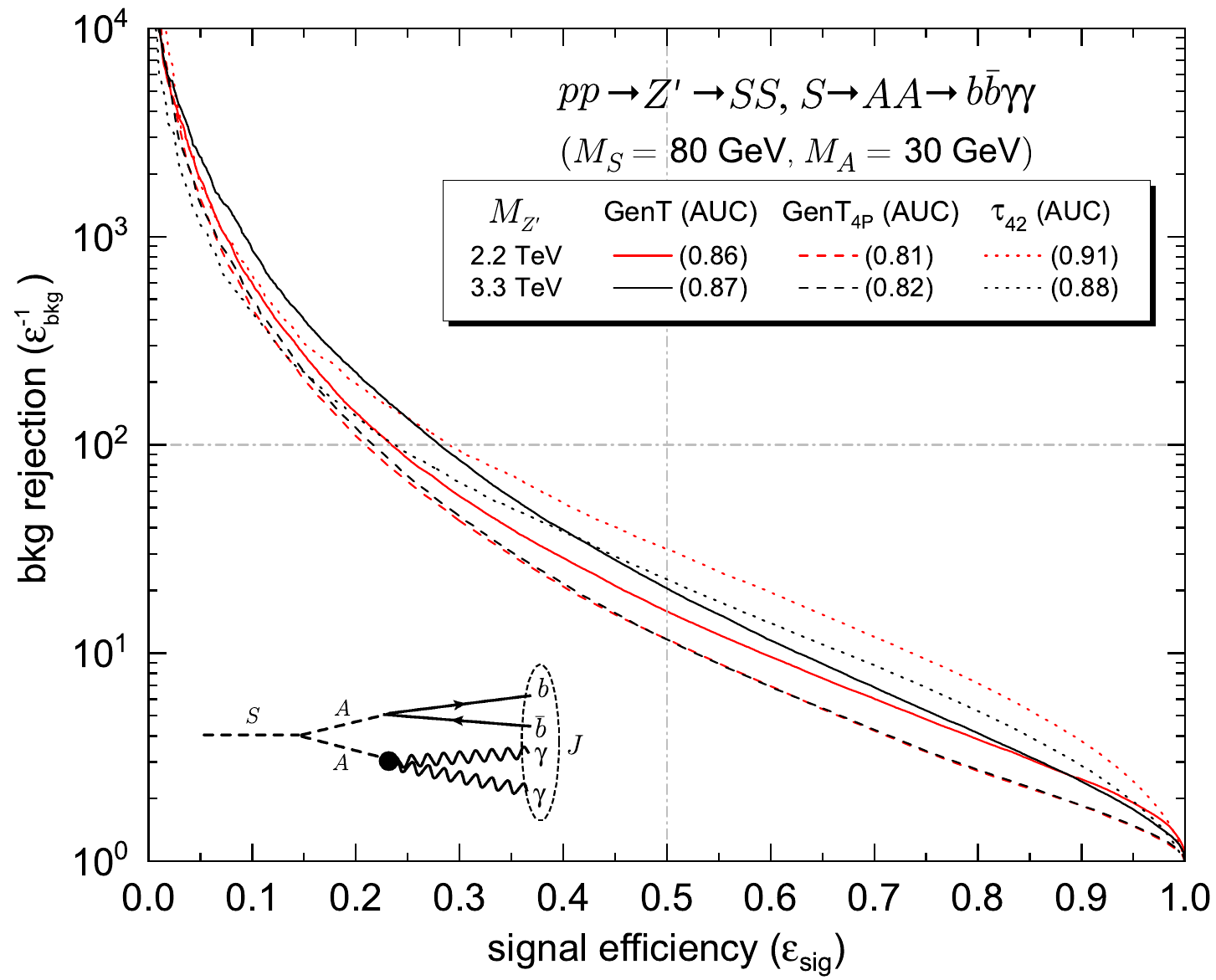} 
 & \includegraphics[height=5.5cm]{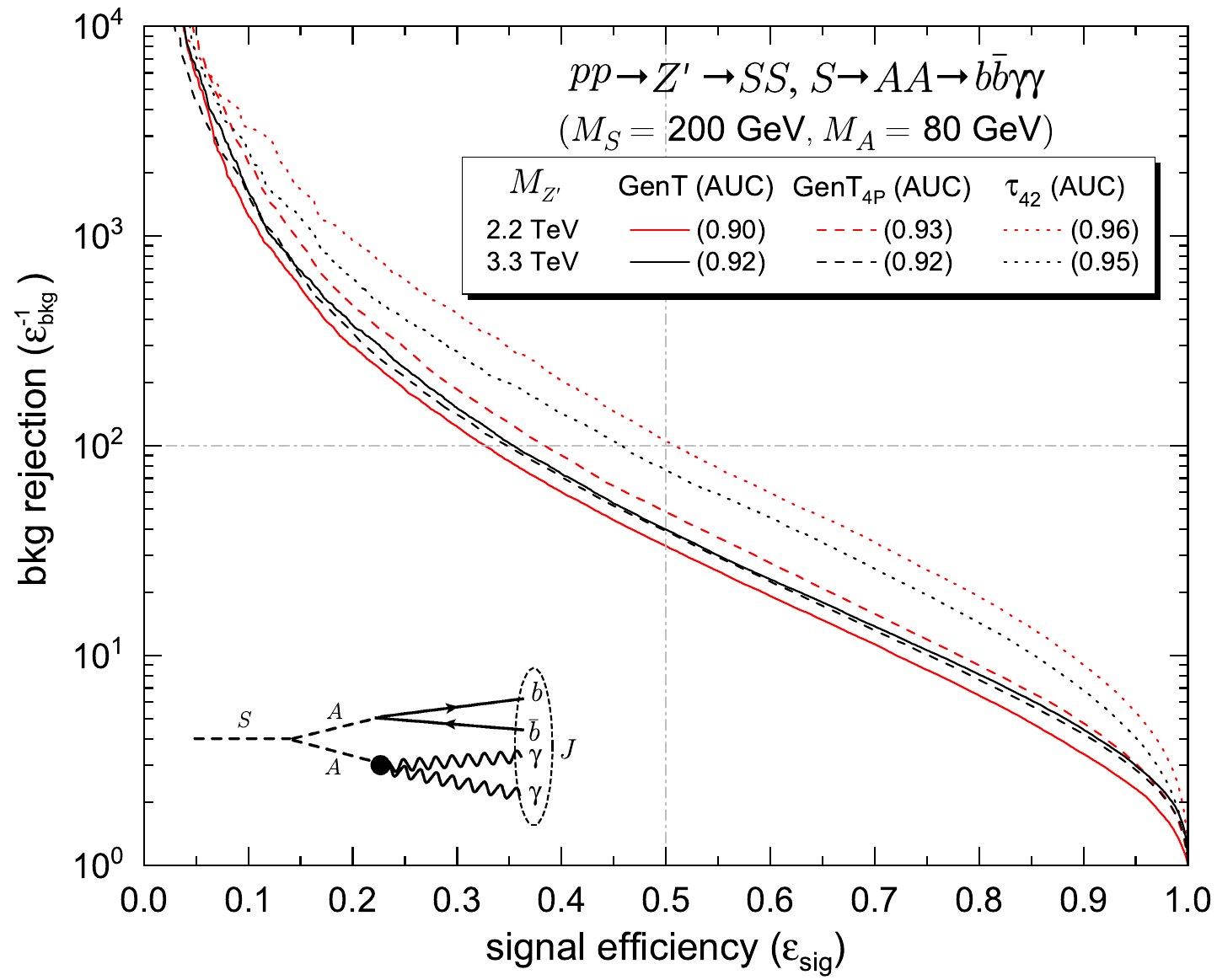} 
\end{tabular}
\caption{ROC curves for other complex jet signals (for details see the self-explanatory legends and the main text).}
\label{Fig:xP}
\end{center}
\end{figure}
In the top panels of Fig.~\ref{Fig:xP}, we present the ROC curves for {\tt GenT} and {\tt GenT$_\text{3P}$}, as well as the ratio $\tau_{32}$, for neutrino jets. Both {\tt GenT} and {\tt GenT$_\text{3P}$} perform quite well for this type of signals, and their inclusion in experimental searches would improve the sensitivity to new physics. As a side comment, in this case the AUC is not a proper measure of the tagger performance, since it is dominated by the region with $\varepsilon_\text{sig} \sim 1$ where the ratio $\tau_{32}$ performs marginally better. On the bottom panels we show the results for {\tt GenT} and {\tt GenT$_\text{4P}$} for jets with hard photons, including also for comparison the ratio $\tau_{42}$, which performs quite well for this type of complex jets (but not for true four-pronged jets, as observed in Fig.~\ref{Fig:4P}). 

These results deserve some discussion. As we have stressed above, the goal of this  analysis was to test whether the taggers are able to identify other types of complex jets for which they are not designed. Or if, on the contrary, they are classified as background-like. The results presented show that our taggers can indeed detect these `unseen' signals with a good efficiency. Note, however, that simpler multivariate methods like a logistic regression may have a better performance~\cite{Aguilar-Saavedra:2020sxp} and taggers including these types of complex jets in the training would provide even better discrimination from the background. Furthermore, for neutrino jets the presence of energetic leptons can be further used for background rejection, as already proposed for top quarks~\cite{Aguilar-Saavedra:2014iga}, and likewise for jets with energetic photons~\cite{AguilarSaavedra:2020wmg}.

\section{Identification of new physics signals}
\label{sec:6}

Let us assume that a new physics signal is discovered, involving a boosted particle with multi-pronged jet signature. There are some cases in which the identity of this boosted particle can be easily established by its leptonic decays, e.g. for $W$ bosons and top quarks. As aforementioned, one can also gain insight on the identity of a new particle by the presence of energetic leptons or photons inside the jet. The most difficult case for discrimination then arises for purely hadronic decays yielding multi-pronged jets.

In order to pinpoint the type of particle that produces such jets, one can use a tagger that identifies their prongness. We build such a tool which takes as input the $N$-subjettiness variables in (\ref{ec:taulist}), using the 2P, 3P and 4P jet samples used for the training of {\tt GenT}. We do not include QCD background jets for simplicity since, from the results in section~\ref{sec:3}, it is clear that multi-pronged jets can already be separated very well from the background. The NN contains two hidden layers of 2048 and 128 nodes, with ReLU activation for the hidden layers and Softmax activation for the three output layers corresponding to the three classes (2P, 3P and 4P). The NN optimisation is performed using the categorical cross-entropy loss function, with the Adam optimiser. 

For each jet in the test samples, the output of the NN provides the relative probabilities $P_{\rm 2P}$, $P_{\rm 3P}$, $P_{\rm 4P}$ that it corresponds to each of the three classes, that is, the probabilities that the jet is two-, three- or four-pronged. The jet is naturally assigned to the class with highest probability $P_\text{sel} = \text{max}(P_{\rm 2P}, P_{\rm 3P}, P_{\rm 4P})$. One can improve the accuracy of the identification by the additional requirement that $P_\text{sel}$ is larger than some threshold value $P_\text{min}$, for example with $P_\text{min} = 0.5$. The downside of this extra constraint is the fact that some of the jets remain {\it undefined}, when neither of the three probabilities reach $P_\text{min}$. 

\begin{figure}[t]
\begin{center}
\includegraphics[height=4.2cm]{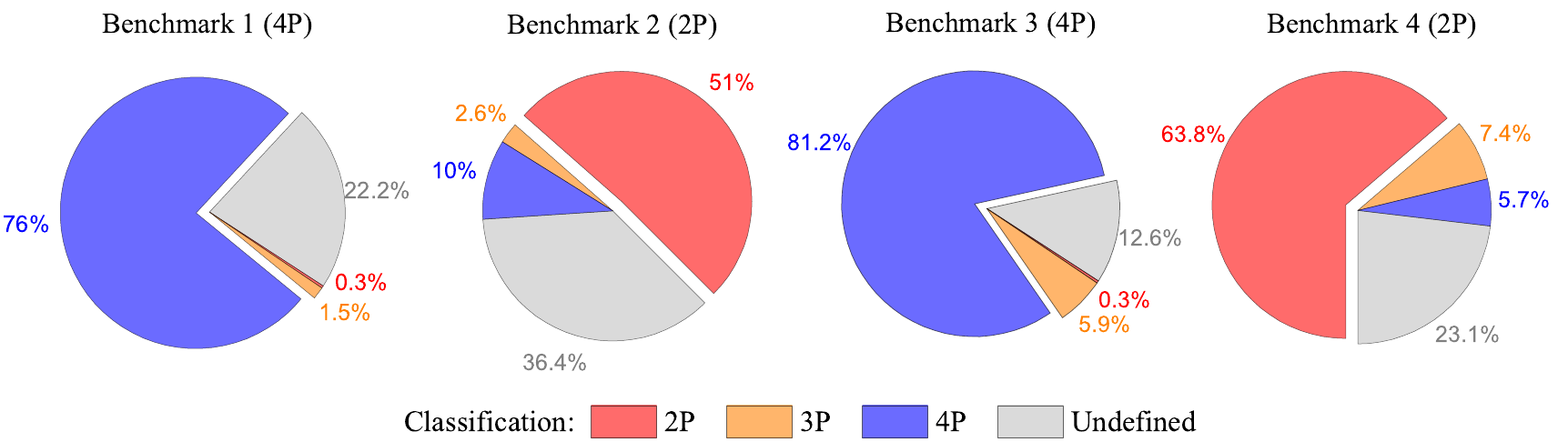} 
\caption{Fraction of jets that are classified as 2P, 3P and 4P, or remain undefined, for the four benchmark examples described in the text.}
\label{Fig:disc}
\end{center}
\end{figure}

Let us study four benchmark examples,
\begin{enumerate}
\item A boosted 80 GeV particle producing four-pronged jets: $Z' \to SS$, $S \to AA \to b \bar b b \bar b$, with $M_{Z'} = 2.2$ TeV, $M_S = 80$ GeV, $M_A = 30$ GeV.
\item  A boosted 80 GeV particle producing two-pronged jets: $Z' \to AA$, $A \to b \bar b$, with $M_{Z'} = 2.2$ TeV, $M_A = 80$ GeV.
\item A boosted 200 GeV particle producing four-pronged jets: $Z' \to SS$, $S \to WW \to q \bar q q \bar q$, with $M_{Z'} = 3.3$ TeV, $M_S = 200$ GeV.
\item  A boosted 200 GeV particle producing two-pronged jets: $Z' \to AA$, $A \to u \bar u$, with $M_{Z'} = 3.3$ TeV, $M_A = 200$ GeV.
\end{enumerate}
In benchmarks (1) and (2) we have 80 GeV jets containing $b$ quarks, whereas in benchmarks (3) and (4) we have 200 GeV jets with light quarks. The relative fraction of jets that is classified as 2P, 3P and 4P, as well as the fraction of jets that remains undefined, is shown for these four benchmarks in Fig.~\ref{Fig:disc}. The identification of four-pronged jets is excellent, with an overwhelming fraction of jets correctly assigned and a minor fraction of jets for which $P_\text{sel} \geq P_\text{min}$ is not verified. The identification of two-pronged jets is quite good as well. Despite the larger percentage of jets that remain undefined, the fraction of correctly identified jets is several times larger than that of misidentified ones. We have checked that in all cases the mistag rates can be further reduced by raising the value of $P_\text{min}$.

\section{Concluding remarks}
In this work we have introduced the novel method of mass unspecific supervised tagging (MUST) for multi-pronged jets. This method avoids the need to build multiple taggers for different jet mass and $p_T$ intervals. The taggers built upon MUST keep an excellent performance across a very wide jet mass and $p_T$ range. In particular, a {\em single} generic tagger {\tt GenT} is able to simultaneously discriminate 2P, 3P and 4P jets from the background across all the jet mass and $p_T$ range. Mass decorrelation, which is an issue for both supervised and non-supervised tools, can easily be implemented by the varying-threshold method. 

In addition, we have successfully verified the performance of our taggers on signals for which they have not been trained, namely jets from heavy neutrinos (including a hard electron) and jets with energetic photons. This broad tagger sensitivity is quite a desirable bonus. Since the possible manifestations of new physics at colliders are yet unknown, tools with a wide scope of usability --- even sensitive to signals not used in the training --- are most valuable. And, of course, these and other types of complex jets may also be included in the training too. Overall, the excellent discrimination power (which often increases with jet $p_T$) and the simplicity of their implementation, make our taggers ideal for the exploration of multi-TeV scales in a wide variety of LHC searches that rely on jet tagging. 

Finally, in the best-case scenario that a new signal is found, uncovering its origin is a must. In this sense, we have developed a selection tagger which is able to determine the prongness of signal jets (2P, 3P, 4P) with a large likelihood. It is unlikely that such discrimination tasks may be performed using unsupervised methods. Therefore, in this respect, the MUST concept for jet tagging could (and hopefully will) play a leading role.

\section*{Acknowledgements}
The work of J.A.A.S. has been supported by MICINN project PID2019-110058GB-C21. F.R.J. and J.F.S. acknowledge Funda\c{c}{\~a}o para a Ci{\^e}ncia e a Tecnologia (FCT, Portugal) for financial support through the projects UIDB/00777/2020, UIDP/00777/2020, CERN/FIS-PAR/0004/2019, and PTDC/FIS-PAR/29436/2017. The work of J.F.S. is supported by the FCT grant SFRH/BD/143891/2019.

\end{document}